\documentstyle[12pt]{article}
\begin{document}

\vspace{10 mm}
\begin{center}
\large{{\bf The Pioneer's acceleration anomaly 
and Hubble's constant.}}

\vspace{3 mm}
\hspace{2 cm}  J.L. Rosales \footnote{
Xerox Corp. The Document Company S.A.U. Ribera del Loira 16-18, E-28042  Madrid (Spain).
E-mail: JoseLuis.Rosales@esp.xerox.com}

\vspace{4 mm}

\end{center}

\vspace{3 mm}

\begin{quote}
\begin{center}
				Abstract
\end{center}

The reported  anomalous acceleration
acting on the Pioneers spacecrafts could  be seen as a consequence of the existence of
some local curvature in light geodesics when using the coordinate speed of
light in an expanding space-time. 
We will show that, as a matter of fact, what has been detected
in the experiments is just the cosmic expansion rate
-the Hubble parameter $H$-, a feature of  light signals. A
relation for the reported annual term is obtained which depends on orbital  
parameters, leading  to suggesting
an analogy between the effect  and Foucault's 
experiment, light rays playing a similar r\^ole 
in the expanding space than  Foucault's Pendulum 
does while determining Earth's rotation.

\end{quote}
\vspace{2 mm}

{\it Pioneer effect.} A careful analysis of orbital data from  Pioneer
10/11 spacecrafts has been reported\cite{kn:Pioneer1},\cite{kn:Pioneer2},\cite{kn:Pioneer3} 
which indicates
the existence  of a very weak acceleration - approximately
$a_p\simeq 8.5 \cdot 10^{-8}cm/s^2$ - apparently directed
toward the Sun. The most conservative (or less adventurous)
hypothesis is that these sorts of  effects on the Pioneers's tracking data
do not entail new
physics and that the detected misfit must be due to  some 
sophisticated (technological) reason 
having to do with  the spacecraft configuration.
However, the analysis in\cite{kn:Pioneer1}-\cite{kn:Pioneer3} seems to have ruled out many
(perhaps all) of such technical
reasons  and the authors even claim  having 
taken into account the accepted
values of the errors in the planetary ephemeris, Earth's orientation, precession,
and nutation.

Thus, in principle, a new effect seems to have unexpectedly 
entered the  phenomenology of physics. 
On the other hand, if such an effect really exists
(i.e., it can not be eliminated by  data reanalysis) it would represent
a violation of Birkhoff's theorem in general relativity for no
constant acceleration at all is  predicted by the Schwarzschild solution.
One of the surprising features of the effect, 
as already pointed out in \cite{kn:Pioneer1}, is that it does not
affect the planets (since no cumulative precession is observed in 
their trajectories) but only to  
spacecrafts (an apparently, strong violation of the equivalence principle.)

 Position determinations procedures for planets and spacecrafts, however, differ.
 In Pioneer's experiment the procedure has to do with transmissions and receptions
 of light signals  between an observer on Earth
 and the object whose coordinate position is to be determined, while for planets
 orbit perturbation theory  is used to obtain
 cumulative result on the full trajectory. That is why Pioneer's coordinate
 determination allows not only to analyze the Solar
 System gravitational field in which the probes are certainly immersed,
 but also the nature of the  non static metric background
 of the expanding space-time on which this gravitational field is acting on
 light itself.

 We will show in this essay that the  nature of the effect 
 lays  on the measurable properties of light rays that propagates within  
 the expanding space background and that what has been detected
 in the experiments is just the cosmic expansion rate
 -the Hubble parameter $H$-, a feature of  light signals.
 The results have, then, nothing to do with the existence
 of dark matter or any other source of local gravitational field.

The full effect is more conveniently understood
when we notice that it deals properly on three different effects: 
\begin{itemize}
 
\item The determination
of an anomalous acceleration acting on the orbit, $a_p$, apparently directed 
toward the Sun 
(although the authors of \cite{kn:Pioneer3} admit that the data
are not provided of
enough resolution as to determine whether the acceleration is 
either toward the Sun or toward the observer on Earth.) 
Moreover, $a_p$ also has a small annual variation amplitude
of the order of $\delta a_p\sim a_p\frac{1}{30}$. 

\item The anomalous Doppler effect which obtains a frequency drift acceleration, $a_t$. 
This is related with $a_p$ upon assuming
$a_t\equiv a_p/c$ 

\item  A  new kind of clock acceleration, $a_q$,
 related but not directly
correlated with the previous $a_t$, which
can better be understood if
there exists an expanding space-time. $a_q$ corresponds to the
detection (see \cite{kn:Pioneer3}) of the  predicted effect  in 
\cite{kn:Rosales-SanchezGomez} that the observed
time of travel  for  light signals  in an expanding space
has a shift of the kind

\begin{equation}
t\rightarrow t-\frac{1}{2}a_q t^2 \mbox{\hspace{2 mm},}
\end{equation}
where $a_q=H$ is Hubble's constant.
\end{itemize}
We will now  derive these three effects from the single 
assumption that there exist a expanding space 
background.

Let us start by considering a generic FRW metric
\begin{equation}
ds^2=-c^2dt^2+\chi(t)^2 dr^2 \mbox{\hspace{2 mm},}
\end{equation}
taking our units of space and time at the cosmological time $t_1$,
we can write,
\begin{equation}
\chi(t)\simeq (\frac{t}{t_1})^{1-\delta} \mbox{\hspace{2 mm},}
\end{equation}
where $\delta \ll 1 $  is a constant depending on the density of the universe, 
and $t_1$ is the local "cosmic time".

Light geodesics satisfy
\begin{equation}
dl\equiv cdt=\chi dr            \mbox{\hspace{2 mm},}
\end{equation}
where $dl$ is the length on the null cone. 

On the other hand, 
we should be able to write our physical laws in such a way that the expansion
of space time be scaled out. This requires using the radial function,
\begin{equation}
r_*\equiv \chi r \mbox{\hspace{2 mm},}
\end{equation}
the metric then becoming
\begin{equation}
ds^2=-(1-\frac{r_*^2H^2}{c^2})c^2dt^2+dr_*^2-2r_*Hdr_*dt \mbox{\hspace{2 mm}.}
\end{equation}
where the local Hubble parameter is
\begin{equation}
H=\frac{d}{dt}\log(\chi) \mbox{\hspace{2 mm}.}
\end{equation}
Since these are not synchronous coordinates (for $g_{0r_*}\neq 0$), 
we define the radial vector
\begin{equation}
\vec{g}=-\frac{r_* H/c}{1-r_*^2H^2/c^2}\vec{r}_1 \mbox{\hspace{2 mm},}
\end{equation}
so that the  space like  element, as measured by some local observer, 
is  the embedded three dimensional metric within the global space time in 
this manifold (see e.g. \cite{kn:Lichnerowicz}) 
\begin{equation}
dl_*^2\equiv\gamma_{r_{*}r_{*}}dr_{*}^2=(g_{r_* r_*}-g_{00}g^2)dr_*^2=\frac{dr_*^2}{1-r_*^2H^2/c^2} \mbox{\hspace{2 mm}.}
\end{equation}

Me may now compare the length, $l_*$ as measured in the locally scaled coordinates, with 
that on the light cone.
In order to do this, notice that one might also have obtained, after (4) and (5), 
the following equation  for the null cone,
\begin{equation}
dl=dr_*(l_*)-r_*(l_*)H\frac{dl}{c} \mbox{\hspace{2 mm},}
\end{equation}
whose solution - using (9), and noting that $\dot{H}\sim O(H^2)$- is
\begin{equation}
l=\frac{c}{H}\log\{1+\sin(\frac{Hl_*}{c})\}\simeq l_*-\frac{Hc}{2}(\frac{l_*}{c})^2+O(H^2) \mbox{\hspace{2 mm}.}
\end{equation}

Which is Equation (1) and represents the measure of the space time curvature on the 
local null cone.

On the light cone,  an effective  
quadratic in time term to the light time must be added, thus mimicking a line of 
sight acceleration of the spacecraft. The constant value  of the
effective residual acceleration  directed
toward the center of coordinates is
\begin{equation}
\kappa=Hc \mbox{\hspace{2 mm}.}
\end{equation}
($l=l_*-\frac{1}{2} \kappa (l_*/c)^2$.) This is the   acceleration observed in Pioneer 10/11
spacecrafts. From the value reported in \cite{kn:Pioneer1} we get a  value for 
the Hubble parameter
\begin{equation}
H=\frac{\kappa}{c}\simeq 83 km/ s \cdot Mpc \mbox{\hspace{2 mm}.}
\end{equation}

The result  depends on the non synchronous character of 
the used coordinates in Equation (6). Thus, the measured  difference between the 
values of the universal time corresponding to two simultaneous events that
take place in two different points of space may be computed as\cite{kn:Landau}
\begin{equation}
\delta t=\frac{1}{c}\int_A^{B} \vec{g}\cdot d\vec{r} \mbox{\hspace{2 mm}.} 
\end{equation}
Which is {\em path dependent}.

This gives, for the spacecrafts, 
\begin{equation}
\delta t = \frac{1}{c}\int_{0}^{l_*} \vec{g}\cdot d\vec{r}\simeq -\frac{1}{2}H(l_*/c)^2  \mbox{\hspace{2 mm}.} 
\end{equation}
The total measurable time for a light signal reaching
Observer's position is given approximately by
\begin{equation}
t_*\simeq l_*/c-\frac{1}{2}H(l_*/c)^2  \mbox{\hspace{2 mm},} 
\end{equation}

which could  also be interpreted in terms of a measurable length given by 
\begin{equation}
l= c_* t_* \simeq l_* -\frac{Hc}{2}(\frac{l_*}{c})^2+O(H^2) \mbox{\hspace{2 mm},}
\end{equation}
which is just our previous result in Eq. (11). 

Yet, the interpretation of our result in terms of the very existence of
an asynchronism in the universal time coordinate 
(depending on the path) between points on  the expanding manifold, also allows
for understanding  that closed trajectories  could not be used to detect
the expansion of the space time; for, if we try to synchronize clocks using
a closed path, we obtain
\begin{equation}
\delta t=\frac{1}{c}\oint \vec{g}\cdot d\vec{r} = 0 \mbox{\hspace{2 mm}.} 
\end{equation}
Which explains why {\em Pioneer effect} leads to no cumulative
measurable effects on the orbits of the planets.
\vspace{6 mm}

Now, let us obtain the relation between $a_p$ and $a_t$. 
For a monochromatic light ray of wave number $k$ and frequency $\omega$ in the metric (6), one easily obtains (see  
e.g. \cite{kn:Landau} \footnote {notice that we use different sign convention in (6) 
than in ref. \cite{kn:Landau}} ),  
\begin{equation}
k_{r_*}=  \frac{\omega}{c}\{\frac{\gamma_{r_{*}r_{*}}}{[-g_{00}]^{1/2}} \frac{dr_{*}}{dl_*} 
+g_{r_*}\} \mbox{\hspace{2 mm} ,}
\end{equation}
which reads as
\begin{equation}
\omega= ck(1+\frac{Hr_*}{c})+ O(H^2) \mbox{\hspace{2 mm} .}
\end{equation}

Correspondingly,
\begin{equation}
\dot{k}=-\frac{\partial \omega}{\partial r_*}=-k H \mbox{\hspace{2 mm},}
\end{equation}
or
\begin{equation}
	k=k_0(1-Ht) + O(H^2) \mbox{\hspace{2 mm},}
\end{equation}
and,
\begin{equation}
\dot{r_*}=\frac{\partial \omega}{\partial k}= c(1+\frac{Hr_*}{c}) \mbox{\hspace{2 mm};}
\end{equation}
finally, from Equations (20) and (22) at the observer's position ($r_*=0$), we obtain
\begin{equation}
\omega =\omega_0 (1-Ht) \mbox{\hspace{2 mm},}
\end{equation}
where $\omega_0= k_0c$. This corresponds to the measured $a_t$
(see \cite{kn:Pioneer3}), i.e.,
\begin{equation}
H=a_t = 2.8 \cdot 10^{-18} s^{-1} \mbox{\hspace{2 mm}.}
\end{equation}

Moreover, from Equation (23), an observer might infer
the presence of an effective Postnewtonian gravitational potential 
\begin{equation}
\phi_H =Hcr_* \mbox{\hspace{2 mm},}
\end{equation}
where $r_*$ is {\it the  coordinate position of the emitter with 
respect to that observer}.
This leads to an apparent vector acceleration that must be added to the orbital
major forces 
\begin{equation}
\vec{a}=-\frac{\partial}{\partial\vec{r}_*}\phi_H = -Hc \frac{\vec{r_*}}{r_*} \mbox{\hspace{2 mm}.}
\end{equation}

Let us now determine $\vec{a}_p$, i.e., that part of the acceleration directed 
toward the Sun. \footnote {the remaining component  
has no effects since its averaged vector value is  
always zero in a year.}
Since $\vec{r}_*=\vec{R}_* -\vec{R}_E $,
where $\vec{R}_E$, and $\vec{R}_*$, are Earth's heliocentric  coordinate vector and 
Pioneer's, one obtains for large  $R_*$
\begin{equation}
\vec{a}_p \approx -Hc(1+\zeta(t)\cos\gamma_0) \frac{\vec{R}_*}{R_*} \mbox{\hspace{2 mm},}
\end{equation}
$\zeta(t)=(R_E/R_*) \cos(2\pi t/T)$.
Here $t=n T$, represents conjunctions and $\gamma_0$,
is the  spacecraft ecliptic latitude  as measured from the Sun. 

The  periodic term pretty explains   
the observation of an additional Doppler drift, correlated with Earth's 
orbital parameters \cite{kn:Pioneer2},\cite{kn:Pioneer3}
\begin{equation}
\delta a_p \approx -Hc \zeta(t)\cos\gamma_0 \mbox{\hspace{2 mm},}
\end{equation} 
which has maxima on conjunctions as reported in \cite{kn:Pioneer2} and \cite{kn:Pioneer3}.
The corresponding total drift in one year period is
\begin{equation}
\frac{\Delta \omega}{\omega}=2 H T \frac{R_E}{R_*}  \cos\gamma_0 \mbox{\hspace{ 2 mm},}
\end{equation}
for Pioneer 10  we can take $< \frac{R_E}{R_*} \cos\gamma_0>_{orbit}\sim 1/30 $ and  
the annual Doppler amplitude becomes
\begin{equation}
<\frac{\Delta \omega}{\omega}>_{orbit}\sim 5\cdot 10^{-12} 
\mbox{ \hspace{ 2 mm} ,}
\end{equation}
the value of the observed annual Doppler amplitude anomaly  coincides with our estimates 
in  Equation (31)\cite{kn:Pioneer3}.
Moreover, Equation (29) also explains the fact that at early times of the 
experiment, the annual term in the anomalous acceleration is largest (see Fig.17 in 
\cite{kn:Pioneer3}.)

Notice the remarkable character of our result in Equations (27) - (30). 
The Pioneer's annual effect depends on the cosine of the ecliptic latitude $\gamma_0$, 
which demonstrates that the annual term is a pure geometrically driven effect.
This  statement could be seen as a benchmark for future experiments.
It also suggest an heuristic analogy with Foucault's experiment;
there, the spin
frequency depends on Pendulum's
colatitude $\lambda$ in a way entirely similar to Equation (30)- recall 
$\omega_{Foucault}=\omega_{Earth}\cos{\lambda}$.

\vspace{4 mm}
We will finally illustrate  the statement that the Pioneer effect is a 
property of light signals, i.e., there exist no additional gravitational field 
apart from  the standard Newtonian one. In order to show this, let us introduce the following 
transformation in the coordinates of Equation (2) for $\delta\simeq 0$. It 
relates the Lorentzian metric
with the Milne ($\chi=Ht$) space time:

\begin{equation}
dt=\frac{1}{(c^2\tau^2-\tilde{r}^2)^{1/2}}[c\tau d\tau -\frac{\tilde{r}}{c}d\tilde{r}] \mbox{\hspace{2 mm},}
\end{equation}
\begin{equation}
dr=\frac{c^2}{H(c^2\tau^2-\tilde{r}^2)}[\tau d\tilde{r}-\tilde{r}d\tau] \mbox{\hspace{2 mm}.}
\end{equation}
Now, $ds^2=0$ leads to
\begin{equation}
\frac{d\tilde{r}}{d\tau}\simeq c \mbox{\hspace{2 mm},}
\end{equation}
i.e.,  $\tilde{r}=c\tau +\tilde{r}_0 $ - independently of the values of $H$.
This means that the effect can be exactly removed in these coordinates.
This satisfactory fact agrees with the consequences of Birkhoff's theorem.

\vspace{6 mm}
{\it Acknowledgements.}   Thanks to the staff of
Xerox Corporation (Spain). I also wish to thank to Dr.Jos\'e Luis S\'anchez-G\'omez from 
the Universidad Aut\'onoma de Madrid for stimulating discussions.

\end{document}